# Several Results in Black Hole Nonformation

Miguel Piñol Ribas, Ignacio López Aylagas

## Abstract

**Black Hole Nonformation was suggested in 2006 by Vachaspati, Stojkovic and Krauss as a sensible solution to the paradox of information, as deduced from calculations in the semiclassical theory of Gravitation. Our own results in the context of classical General Relativity (with the addition of several thermodynamic argumentations) led us to an equivalent statement: collapsing bodies emit their complete amount of mass and energy as radiation before the formation of any event horizon. In this paper, we submit the metric we rendered to the test of Einstein Field Equations, and we propose some modifications to extend our solutions to more general situations, as the initial moments of collapse and electrically charged collapses, as well as we effectuate some considerations about rotating collapses.**

Contact e-mail: mpinolri7@alumnes.ub.edu

## Introduction

Black Holes hold a crucial position in modern Theoretical Physics, and the major part of the scientific community admits them as an undisputed truth. Nonetheless, a number of authors (including the celebrated physicist Marcel Brillouin) have pointed out several logical objections to the coherence of their concept as it is usually understood [1]. Unfortunately, all them seem to have been systematically ignored.

Among black hole criticizers, most of them have considered Hilbert's space-time to constitute a grievous misunderstanding of Schwarzschild's solution, and alternative vacuum solutions of Einstein's equations have been developed. For instance, Loinger and Marisco [2] remark that the metric actually due to Schwarzshild is this one:

$$ds^2_{Schw} = -\left(1 - \frac{2M}{f(r)}\right)dt^2 + \left(1 - \frac{2M}{f(r)}\right)^{-1}[d f(r)]^2 + [f(r)]^2 d\Omega^2 \quad \text{(eq. 1)},$$

with $f(r)$ being any regular function in $r$, while the concretion $f(r) = r$ really corresponds to Hilbert:

$$ds^2_{Hilbert} = -\left(1 - \frac{2M}{r}\right)dt^2 + \left(1 - \frac{2M}{r}\right)^{-1}dr^2 + r^2 d\Omega^2 \quad \text{(eq. 2)}$$

Even so, in this paper, without entering the historical controversy, from now on we will use the usual name of "Schwarzshild metric" for the metric specified by equation "2": not in order to assert it was actually supplied by Schwarzshild, but since the simple reason that it is widely known by that denomination.

As a matter of fact, our criticism to black holes does not rise from considering equation "2" a bad punctual-mass solution, but from the mathematical constatation that the structure of the metric itself constitutes an impediment to the formation of a punctual-mass by a process of collapse: in truth we consider that a process of collapse is much better described by a never-stationary metric, and we rendered a metric of such style in our previous work [3].

Black Hole Nonformation had been already suggested in 2006 by Vachaspati, Stojkovic and Krauss [4], who used the methods of semiclassical theory in order to calculate the quantum radiation emitted by a collapsing shell approaching an event horizon, and the result was obtained that the collapsing shell should be fully evaporated before having reached the event horizon. Their result is highly remarkable, as it holds a clear solution to the paradox of information. In 2008, Vachaspati and Stojkovic reviewed their calculations by dealing simultaneously the infalling matter and the radiation field both in quantum theory, and not only the radiation field as in semiclassical theory, but the same behaviour was observed [5].

The metric that we rendered (Never-Stationary Gravitational Collapse metric), deduced as a good approximation for large times, may be defined by the following formulae:

$$ds^2_{Scw-P} = -h_{(r,t)} dt^2 + h_{(r,t)}^{-1} dr^2 + r^2 d\Omega^2 \quad \text{(eq. 3a)}$$

$$h_{(r,t)} = h_{(r,t)}^{outside} H(r - r_{edge}) + h_{(r,t)}^{inside} H(r_{edge} - r) \quad \text{(eq. 3b)}$$

$$h_{(r,t)}^{outside} = 1 - \frac{2M}{r} \quad \text{(eq. 3c)}$$

$$h_{(r,t)}^{inside} = e^{f(r) - \frac{t}{r}} \quad \text{(eq. 3d)}$$

$$r_{edge} = 2M \left[ 1 + e^{f(2M) - \frac{t}{2M}} \right] \quad \text{(eq. 3e)}$$

We also proposed an "extended form" of the Never-Stationary Gravitational Collapse metric, which we thought to be useful for further studies and the research of a higher degree of precision, by including a factor $\beta_{(r,t)}$ in order to have into account that the speed of collapse must be only a fraction of the speed of light:

$$h_{(r,t)}^{inside} = \left[ e^{f(r)} - \frac{2}{r} P_{(r/2)} \right] e^{\frac{-1}{r} \int \beta dt} + \frac{2}{r} P_{(r/2)} \quad \text{(eq. 4)}$$

It may be noticed that we added as well some "radiative corrections", $P_{(\mu)}$, due to thermodynamic grounds. If we neglect them, we get the following expression:

$$h_{(r,t)}^{inside} = e^{f(r) - \frac{1}{r} \int \beta \, dt} \quad \text{(eq. 5)}$$

In this paper, we carefully study several aspects related to the Never-Stationary Gravitational Collapse metric. First, we detailedly check if the Never-Stationary Gravitational Collapse metric accomplishes the Einstein Field Equations. In second place, a model for initial collapse is presented. Third, we apply to the Reissner-Nördstrom metric an equivalent treatment to that which we applied to the Schwarzschild metric in order to convert it into the Never-Stationary Gravitational Collapse metric. Finally, some considerations about symmetry in respect to temporal inversion lead us to assert that the restriction of black hole formation must be extended as well to rotating collapses.

RESULTS

**A) The Never-Stationary Gravitational Collapse metric accomplishes the Einstein Field Equations**

The Never-Stationary Gravitational Collapse metric presents a structure of the following type:

$$ds^2 = -e^{\nu} dt^2 + e^{\lambda} dr^2 + r^2 d\Omega^2 \quad \text{(eq. A1)}$$

Among this family of metrics, the Never-Stationary Gravitational Collapse metric (in its extended form) is concretely defined by the following functions $\nu_{(r,t)}$ and $\lambda_{(r,t)}$:

$$\nu_{(r,t)} = -\lambda_{(r,t)} = \ln\left(1 - \frac{2\mu_{(r,t)}}{r}\right) = f(r) - \frac{1}{r} \int \beta_{(r,t)} dt \sim f(r) - \frac{t}{r} \quad \text{(eq. A2)},$$

where $\mu_{(r,t)}$ is the total energy-mass contained inside a surface of the radius $r$ at time $t$, and $\beta_{(r,t)}$ is the quotient between the speed of collapse and the speed of light.

The non-trivial Einstein Field Equations for a metric of the type detailed in equation "A1" are those specified in here [6]:

$$8\pi T_0^0 = -e^{-\lambda}\left(\frac{1}{r^2} - \frac{\lambda'}{r}\right) + \frac{1}{r^2} \quad \text{(eq. A3a)}$$

$$8\pi T_0^1 = -e^{-\lambda}\frac{\dot{\lambda}}{r} \quad \text{(eq. A3b)}$$

$$8\pi T_1^1 = -e^{-\lambda}\left(\frac{\nu'}{r} + \frac{1}{r^2}\right) + \frac{1}{r^2} \quad \text{(eq. A3c)}$$

$$8\pi T_2^2 = -\frac{1}{2}e^{-\lambda}\left(\nu'' + \frac{\nu'^2}{2} + \frac{\nu' - \lambda'}{r} - \frac{\nu'\lambda'}{2}\right) + \frac{1}{2}e^{-\nu}\left(\ddot{\lambda} + \frac{\dot{\lambda}^2}{2} - \frac{\dot{\lambda}\dot{\nu}}{2}\right) \quad \text{(eq. A3d)}$$

$$8\pi T_3^3 = 8\pi T_2^2 \quad \text{(eq. A3e)}$$

We have used the prime $(')$ in order to mean differentiation with respect to $r$, and a dot $(\cdot)$ on a symbol means differentiation with respect to $t$.

Let's now analyse if the functions proposed in equation "A2" are in agreement with the Einstein Field Equations "A3a-e".

In first place, we are going to examine equation "A3a". The $T_0^0$ component of the energy-momentum tensor corresponds to the density $\rho_{(r,t)}$ of matter-energy, which we may calculate by considering spherical symmetry in the following way:

$$T_0^0 = \rho_{(r,t)} = \frac{d\mu_{(r,t)}}{dV} = \frac{1}{4\pi r^2}\frac{d\mu_{(r,t)}}{dr} \quad \text{(eq. A4a)},$$

where $dV$ is shell of volume of radius $r$ and width $dr$, so that $dV = 4\pi r^2 dr$.

We can isolate an expression for $\mu_{(r,t)}$, in function of $\lambda_{(r,t)}$, from equation "A2":

$$\mu_{(r,t)} = \frac{r}{2}\left(1 + e^{-\lambda_{(r,t)}}\right) \quad \text{(eq. A4b)}$$

Therefore:

$$T_0^0 = \rho_{(r,t)} = \frac{1}{4\pi r^2}\frac{d}{dr}\left[\frac{r}{2}\left(1 + e^{-\lambda_{(r,t)}}\right)\right] = \frac{1}{4\pi r^2}\left[\frac{1}{2}\left(1 + e^{-\lambda_{(r,t)}}\right) - \frac{r}{2}e^{-\lambda_{(r,t)}}\lambda'\right] \quad \text{(eq. A4c)}$$

By grouping factors and terms, equation "A4c" exactly reproduces equation "A3a", that is what we wanted to prove.

In second place, we are going to study equation "A3b". The $T_0^1$ component of the energy-momentum tensor corresponds to the flux of matter-energy through a surface of constant $x^1 = r$. This flux is equivalent to the product of density $\rho_{(r,t)}$ and "radial speed" $v_{r(r,t)}$, and radial speed is at the same time the product of the "speed of a ray of light moving along the radial direction" $c_{r(r,t)}$ and the quotient $\beta_{(r,t)}$.

$$T_0^1 = \rho_{(r,t)} v_{r(r,t)} = \rho_{(r,t)} \beta_{(r,t)} c_{r(r,t)} \quad \text{(eq. A5a)}$$

The speed of a ray of light moving along the radial direction may be obtained by imposing the conditions $ds^2=0$ (light-like trajectory) and $d\Omega^2=0$ (purely radial movement) to the metric "A1":

$$|c_{r(r,t)}|=e^{\frac{\nu-\lambda}{2}} \quad \text{(eq. A5b)}$$

According to equation "A2", $\nu_{(r,t)}=-\lambda_{(r,t)}$, and we must impose a minus sign to the previous equation in order to have into account that we are dealing with a collapse and not an expansion:

$$c_{r(r,t)}=-e^{-\lambda} \quad \text{(eq. A5c)}$$

Therefore:

$$T^1_0=-e^{-\lambda}\beta_{(r,t)}\rho_{(r,t)} \quad \text{(eq. A5d)}$$

If equation "A2" is correct, at large times function $\lambda_{(r,t)}$ will exhibit very elevated values, and consequently $e^{-\lambda}$ could be ignored when added to much greater terms (but we are not going to ignore it when the consequence would be a null factor). Thus, equations "A4b" and "A4c" may be approximated to the next expressions:

$$\mu_{(r,t)}=\frac{r}{2}+O(e^{-\lambda}) \quad \text{(eq. A5e)}$$

$$\rho_{(r,t)}=\frac{1}{8\pi r^2}+O(e^{-\lambda}) \quad \text{(eq. A5f)}$$

Furthermore, if equation "A2" is correct, the relationship between $\beta_{(r,t)}$ and $\lambda_{(r,t)}$ results to be the following one:

$$\frac{1}{r}\beta_{(r,t)}=\dot{\lambda} \quad \text{(eq. A5g)}$$

By introducing equations "A5f" and "A5e" into equation "A5d", we obtain:

$$T^1_0=-e^{-\lambda}(r\dot{\lambda})\left[\frac{1}{8\pi r^2}+O(e^{-\lambda})\right]=\frac{-1}{8\pi}e^{-\lambda}\frac{\dot{\lambda}}{r}+O(e^{-2\lambda}) \quad \text{(eq. A5h)}$$

By grouping factors, equation "A5h" exactly coincides with equation "A3b", except by the corrective terms depending on $e^{-2\lambda}$. Thus, we may conclude that the Never-Stationary Gravitational Collapse metric defined by equation "A2" constitutes a good solution for gravitational collapse at large times, but not for incipient collapse (as one of the steps in the derivation of the metric implied that every piece of matter came close to the surface where an event horizon would be formed if it actually crossed it).

In spite of all, from equations "A3b", "A4c" and "A5d", a more precise differential equation for $\lambda_{(r,t)}$:

$$\dot{\lambda}=\frac{d\lambda_{(r,t)}}{dt}=r\frac{8\pi T^1_0}{-e^{-\lambda}}=r\frac{8\pi\left(-e^{-\lambda}\beta_{(r,t)}\rho_{(r,t)}\right)}{-e^{-\lambda}}=r\beta_{(r,t)}\left(8\pi\rho_{(r,t)}\right) \quad \text{(eq. A6a)}$$

$$\frac{d\lambda_{(r,t)}}{dt}=r\beta_{(r,t)}\left[\frac{1}{r^2}-e^{-\lambda}\left(\frac{1}{r^2}-\frac{\lambda'}{r}\right)\right]=\frac{\beta_{(r,t)}}{r}\left[1-e^{-\lambda}(1-r\lambda')\right] \quad \text{(eq. A6b)}$$

$$\lambda_{(r,t)}=\frac{1}{r}\int\beta_{(r,t)}\left[1-e^{-\lambda}(1-r\lambda')\right]dt \quad \text{(eq. A6c)}$$

(We are going to study in a much more detailed way the incipient gravitational collapse in the next section.)

Equations "A3c-e" still remain to be analysed, but certainly those are not going to imply a great complication. We are mainly interested in their approximative expressions for large times:

$$8\pi T_1^1 = \frac{1}{r^2} + O(e^{-\lambda}) \quad \text{(eq. A7a)}$$

$$8\pi T_2^2 = 8\pi T_3^3 = \frac{1}{2} e^{-f(r) + \frac{1}{r}\int \beta dt} \left( \frac{\dot{\beta}}{r} + \frac{\beta^2}{r^2} \right) + O(e^{-\lambda}) \quad \text{(eq. A7b)}$$

The diagonal components of the energy-momentum tensor represent the "pressure":

$$T_1^1 = P_r, \; T_2^2 = P_\theta, \; T_3^3 = P_\phi \quad \text{(eq. A7c)}$$

We can check how equations "A6a-b" give positive well-behaved functions for the pressure. $P_r$ tends to a constant value, while $P_\theta$ and $P_\phi$ are continuously growing with time, as the collapse progresses. A suitable equation of state should be necessary in order to perform a deeper study (which would allow a precise determination of $\beta_{(r,t)}$ ).

**B) A model for the incipient collapse**

In order to study the incipient collapse, we are going to consider it as a "free fall". We know that this assumption is not going to reproduce the exact physical situation, for the inner shells will exert a pressure over the outer one, but nevertheless it is also true that the pressure is not expected to be large in the initial moments.

We depart from these equations:

$$ds^2 = -e^{-\lambda} dt^2 + e^{\lambda} dr^2 + r^2 d\Omega^2 \quad \text{(eq. B1a)}$$

$$e^{-\lambda} = 1 - \frac{2\mu}{r} \quad \text{(eq. B1b)}$$

$$r = \frac{2\mu}{1 - e^{-\lambda}} \quad \text{(eq. B1c)},$$

which we have already presented in the previous section.

The equation for the conservation of energy of a free falling particle $p$ in the Schwarzschild metric corresponds to the following one [7]:

$$\left( \frac{dr_p}{d\tau} \right)^2 = \eta^2 - \left( 1 - \frac{2\mu}{r_p} \right)\left( 1 + \frac{l^2}{r_p^2} \right) \quad \text{(eq. B2a)},$$

where $\eta$ is the energy per unit mass of the particle, and $l$ its angular momentum per unit mass. As we are considering a purely radial collapse, $l = 0$ :

$$\left( \frac{dr_p}{d\tau} \right)^2 = \eta^2 - \left( 1 - \frac{2\mu}{r_p} \right) \quad \text{(eq. B2b)}$$

The relationship between the proper time $\tau$ measured by the particle and the time $t$ measured by an asymptotic observer can be deduced from the metric "B1a", by having into account that $ds^2 = d\tau^2$ and that in a purely radial movement $d\Omega^2 = 0$ :

$$d\tau^2 = e^{-\lambda} dt^2 - e^{\lambda} dr_p^2 \quad \text{(eq. B3a)}$$

$$\left(\frac{d\tau}{dt}\right)^2 = e^{-\lambda} - e^{\lambda}\left(\frac{dr_p}{dt}\right)^2 \quad \text{(eq. B3b)}$$

By "multiplying" equation "C3b" and equation "B2b", we obtain:

$$\left(\frac{dr_p}{d\tau}\right)^2 \left(\frac{d\tau}{dt}\right)^2 = \left[\eta^2 - \left(1 - \frac{2\mu}{r_p}\right)\right]\left[e^{-\lambda} - e^{\lambda}\left(\frac{dr_p}{dt}\right)^2\right] \quad \text{(eq. B4a)}$$

$$\left(\frac{dr_p}{dt}\right)^2 = \left[\eta^2 - \left(1 - \frac{2\mu}{r_p}\right)\right]\left[e^{-\lambda} - e^{\lambda}\left(\frac{dr_p}{dt}\right)^2\right] \quad \text{(eq. B4b)}$$

As we have argued in section A, the "speed of radial collapse", $v_r$ (which here coincides with the speed of the free falling particle $\frac{dr_p}{dt}$), may be expressed as a fraction of the speed of light, in the following way:

$$\frac{dr_p}{dt} = v_{r(r,t)} = \beta_{(r,t)} c_{r(r,t)} = -\beta_{(r,t)} e^{-\lambda} \quad \text{(eq. B5)}$$

By replacing equations "B1b" and "B5" into equation "B4b", we find:

$$\left(-\beta_{(r,t)} e^{-\lambda}\right)^2 = \left[\eta^2 - e^{-\lambda}\right]\left[e^{-\lambda} - e^{\lambda}\left(-\beta_{(r,t)} e^{-\lambda}\right)^2\right] \quad \text{(eq. B6a)}$$

$$e^{-2\lambda} \beta^2_{(r,t)} = \left[\eta^2 - e^{-\lambda}\right] e^{-\lambda} \left[1 - \beta^2_{(r,t)}\right] \quad \text{(eq. B6b)}$$

$$e^{-\lambda} \beta^2_{(r,t)} = \left[\eta^2 - e^{-\lambda}\right]\left[1 - \beta^2_{(r,t)}\right] \quad \text{(eq. B6c)}$$

$$\beta^2_{(r,t)} = 1 - \frac{e^{-\lambda}}{\eta^2} \quad \text{(eq. B6d)}$$

$$\beta_{(r,t)} = \pm\sqrt{1 - \frac{e^{-\lambda}}{\eta^2}} \quad \text{(eq. B6e)}$$

As a collapse is being considered and not an expansion, the plus (+) sign must be chosen:

$$\beta_{(r,t)} = \sqrt{1 - \frac{e^{-\lambda}}{\eta^2}} \quad \text{(eq. B6f)}$$

In order to obtain a differential equation, we must ask ourselves "What is the precise relationship between $\beta_{(r,t)}$ and $\dot{\lambda}_{(r,t)}$?" In the previous section, this one was used:

$$\dot{\lambda} = \frac{\beta}{r} \quad \text{(eq. B7a)},$$

but it had been deduced from equation "A2", which was proved to be a good solution for large times, not for incipient collapse (that is precisely what we are wishing to describe now). A more accurate expression may be derived by isolating $\beta_{(r,t)}$ from equation "B5" and inserting equation "B1c" in it:

$$\beta_{(r,t)} = \frac{\frac{dr_p}{dt}}{c_{r(r,t)}} = \frac{\frac{d}{dt}(r_p)}{-e^{-\lambda}} = \frac{\frac{d}{dt}\left(\frac{2\mu}{1-e^{-\lambda}}\right)}{-e^{-\lambda}} = \frac{\frac{+2\mu e^{-\lambda}(-\dot{\lambda})}{(1-e^{-\lambda})^2}}{-e^{-\lambda}} = \frac{2\mu\dot{\lambda}}{(1-e^{-\lambda})^2} \quad \text{(eq. B7b)}$$

It can straightforwardly checked how when $\lambda_{(r,t)}$ takes elevate values, equations "B7a" and "B7b" coincide (attending the relation expressed in equation "B1c").

From equations "B6f" and "B7b", the following differential equation may be deduced:

$$\frac{2\mu\dot{\lambda}}{(1-e^{-\lambda})^2} = \sqrt{1 - \frac{e^{-\lambda}}{\eta^2}} \quad \text{(eq. B8a)}$$

By having into account that $\dot{\lambda} \equiv \frac{d\lambda}{dt}$, a separation of variables can be performed:

$$\frac{dt}{2\mu} = \frac{d\lambda}{(1-e^{-\lambda})^2 \sqrt{1 - \frac{e^{-\lambda}}{\eta^2}}} \quad \text{(eq. B8b)}$$

In the integral form:

$$\int \frac{dt}{2\mu} = \int \frac{d\lambda}{(1-e^{-\lambda})^2 \sqrt{1 - \frac{e^{-\lambda}}{\eta^2}}} \equiv I_t(t) = I_\lambda(\lambda) \quad \text{(eq. B8c)}$$

The first part of equation "B8c" is trivially integrable:

$$I_t(t) = \int \frac{dt}{2\mu} = \frac{t}{2\mu} \quad \text{(eq. B9)}$$

About the second part, in order to solve it the following change of variables may be convenient:

$$\sqrt{1 - \frac{e^{-\lambda}}{\eta^2}} = z \quad \text{(eq. B10a)}$$

Therefore:

$$dz = d\left(\sqrt{1 - \frac{e^{-\lambda}}{\eta^2}}\right) = \frac{\frac{-1}{\eta^2}e^{-\lambda}(-d\lambda)}{2\sqrt{1 - \frac{e^{-\lambda}}{\eta^2}}} = \frac{1-z^2}{2z}d\lambda \quad \text{(eq. B10b)}$$

$$d\lambda = \frac{2z}{1-z^2}dz \quad \text{(eq. B10c)}$$

$$e^{-\lambda} = \eta^2(1-z^2) \quad \text{(eq. B10d)}$$

Thus:

$$\frac{d\lambda}{(1-e^{-\lambda})^2\sqrt{1-\frac{e^{-\lambda}}{\eta^2}}} = \frac{\frac{2z}{1-z^2}dz}{(1-\eta^2+\eta^2 z^2)^2 z} = \frac{\frac{2}{\eta^4}dz}{(1-z^2)\left(z^2+\frac{1-\eta^2}{\eta^2}\right)^2} \quad \text{(eq. B11)}$$

The previous function of $z$ can be divided in several terms, what will help to integrate it:

$$\frac{\frac{2}{\eta^4}}{(1-z^2)\left(z^2+\frac{1-\eta^2}{\eta^2}\right)^2} = \frac{1}{1+z}+\frac{1}{1-z}+\frac{2}{z^2+\frac{1-\eta^2}{\eta^2}}+\frac{\frac{2}{\eta^2}}{\left(z^2+\frac{1-\eta^2}{\eta^2}\right)^2} \quad \text{(eq. B12)}$$

At this point, integration is easy to perform:

$$I_\lambda = \int\left(\frac{1}{1+z}+\frac{1}{1-z}+\frac{2}{z^2+\frac{1-\eta^2}{\eta^2}}+\frac{\frac{2}{\eta^2}}{\left(z^2+\frac{1-\eta^2}{\eta^2}\right)^2}\right)dz \quad \text{(eq. B13a)}$$

$$I_\lambda = \ln(1+z)-\ln(1-z)+A(z,\eta) = \ln\left(\frac{1+z}{1-z}\right)+A(z,\eta) \quad \text{(eq. B13b)}$$

The expression of $A(z,\eta)$ will depend upon the value of $\eta$ being equal, superior or inferior to the unity. As $\eta$ represents the energy per unit mass, and as we are considering collapse as one of the final possible stages of a closed system, we will consider $\eta<1$, and consequently the expression for $A(z,\eta)$ will be the following one:

$$A(z,\eta) = 2a_{(\eta)}\operatorname{arctg}\left(a_{(\eta)}z\right) - \frac{a_{(\eta)}^3}{\eta^2}\left[\frac{2+a_{(\eta)}^2 z^2}{a_{(\eta)}z\left(1+a_{(\eta)}^2 z^2\right)} + \operatorname{arctg}\left(a_{(\eta)}z\right)\right]+C \quad \text{(eq. B14a)},$$

where:

$$a_{(\eta)} \equiv \sqrt{\frac{\eta^2}{1-\eta^2}} \quad \text{(eq. B14b)},$$

and where $C$ is a constant of integration.

Let's study a bit more carefully the first two terms in equation "B13b", in order to make more explicit the relation with $\lambda_{(r,t)}$:

$$I_\lambda - A(z,\eta) = \ln\left(\frac{1+z}{1-z}\right) = \ln\left(\frac{1+z}{1-z}\frac{1+z}{1+z}\right) = \ln\left(\frac{1+2z+z^2}{1-z^2}\right) \quad \text{(eq. B15a)}$$

$$I_\lambda - A(z,\eta) = \ln\left[\frac{1+2\sqrt{1-\frac{e^{-\lambda}}{\eta^2}}+\left(1-\frac{e^{-\lambda}}{\eta^2}\right)}{\frac{e^{-\lambda}}{\eta^2}}\right] = \lambda + \ln\left(2\eta^2+2\eta\sqrt{\eta^2-e^{-\lambda}}-e^{-\lambda}\right) \quad \text{(eq. B15b)}$$

Finally, from equations "B8c", "B9" and "B15b", we can conclude the dependence between time and the coefficient

$\lambda_{(r,t)}$ , which was what we were searching:

$$\frac{t}{2\mu} = \lambda + \ln\left(2\eta^2 + 2\eta\sqrt{\eta^2 - e^{-\lambda}} - e^{-\lambda}\right) + A(z(\lambda)) \quad \text{(eq. B16a)},$$

or, if we replace $2\mu$ by $r = \dfrac{2\mu}{1-e^{-\lambda}}$ :

$$\frac{t}{r} = \left(1-e^{-\lambda}\right)\left[\lambda + \ln\left(2\eta^2 + 2\eta\sqrt{\eta^2 - e^{-\lambda}} - e^{-\lambda}\right) + A(z(\lambda))\right] \quad \text{(eq. 16b)}$$

Two limits of this equation should be studied. The first one corresponds to $t=0$ , that will allow us to provide a value for the constant of integration $C$ in function of the initial parameters of collapse (that is, of the initial massive structure of the star which enters collapse, $2\mu_0(r)$ ):

$$0 = \lambda_0 + \ln\left(2\eta^2 + 2\eta\sqrt{\eta^2 - e^{-\lambda_0}} - e^{-\lambda_0}\right) + A(z(\lambda_0)) \quad \text{(eq. B17a)}$$

$$A(z(\lambda_0)) = 2a_{(\eta)} arctg(a_{(\eta)} z(\lambda_0)) - \frac{a_{(\eta)}^3}{\eta^2}\left[\frac{2 + a_{(\eta)}^2 (z(\lambda_0))^2}{a_{(\eta)} z\left(1 + a_{(\eta)}^2 (z(\lambda_0))^2\right)} + arctg(a_{(\eta)} z(\lambda_0))\right] + C \quad \text{(eq. B17b)}$$

$$z(\lambda_0) = \sqrt{1 - \frac{e^{-\lambda_0}}{\eta^2}} \quad \text{(eq. B17c)}$$

$$\lambda_0 = \lambda_0(r) = -\ln\left(1 - \frac{2\mu_0(r)}{r}\right) \quad \text{(eq. B17d)},$$

$C$ will be constant in time, but it the general case it will be a function of $r$ , $C = C(r)$ .

The other interesting limit consists in the behaviour of the system at large times, when $e^{-\lambda_0} \ll 1$ and $z \approx 1$ :

$$\frac{t}{2\mu} \approx \frac{t}{r} \approx \lambda + \ln(4\eta^2) + A(z=1) \quad \text{(eq. B18a)}$$

$$A(z=1) = 2a_{(\eta)} arctg(a_{(\eta)}) - \frac{a_{(\eta)}^3}{\eta^2}\left[\frac{2 + a_{(\eta)}^2}{a_{(\eta)}(1 + a_{(\eta)}^2)} + arctg(a_{(\eta)})\right] + C(r) \quad \text{(eq. B18b)}$$

Comparing equation "B18a" and equation "A2", the expression of $f(r)$ in terms of the contour conditions of the problem may be determined:

$$f(r) \sim \ln(4\eta^2) + A(z=1) \quad \text{(eq. B19)}$$

**C) Derivation of a metric for a collapsing non-rotating charged body**

The metric which describes a charged non-rotating black hole is not the Schwarzschild metric, but Reissner-Nördstrom metric:

$$ds^2_{RN} = -\left(1 - \frac{2M}{r} + \frac{Q^2}{r^2}\right)dt^2 + \left(1 - \frac{2M}{r} + \frac{Q^2}{r^2}\right)^{-1} dr^2 + r^2 d\Omega^2 \quad \text{(eq. C1)},$$

where $M$ is the total mass of the black hole and $Q$ its charge.

We are going now to consider a process of collapse, where $\mu_{(r,t)}$ and $q_{(r,t)}$ will be functions of $r$ and $t$, and they will respectively correspond to the total mass and charge delimited by a surface of radius $r$ at time $t$. However, in order to study the process, we will consider the motion of a spherical Σ-surface co-mobile with the collapsing energy-matter, so that for the surface $\mu_\Sigma$ and $q_\Sigma$ will be constants.

If the motion is purely radial, with no angular momentum, $d\Omega^2 = 0$. If the process of collapse took place at the speed of light, then we would have $ds^2 = 0$:

$$0 = -\left(1 - \frac{2\mu_\Sigma}{r_{light}} + \frac{q_\Sigma^2}{r_{light}^2}\right) dt^2 + \left(1 - \frac{2\mu_\Sigma}{r_{light}} + \frac{q_\Sigma^2}{r_{light}^2}\right)^{-1} dr_{light}^2 \quad \text{(eq. C2a)}$$

$$\left(\frac{dr_{light}}{dt}\right)^2 = \left(1 - \frac{2\mu_\Sigma}{r_{light}} + \frac{q_\Sigma^2}{r_{light}^2}\right)^2 \quad \text{(eq. C2b)}$$

$$\left|\frac{dr_{light}}{dt}\right| = 1 - \frac{2\mu_\Sigma}{r_{light}} + \frac{q_\Sigma^2}{r_{light}^2} \quad \text{(eq. C2c)}$$

As we are considering a collapse and not an expansion, the sign of the speed must be negative. As the collapse will be taking place at a fraction of the speed of light, and thus a factor $\beta_{(r,t)}$ should be added:

$$\frac{dr_\Sigma}{dt} = -\beta_{(r,t)}\left[1 - \frac{2\mu_\Sigma}{r_\Sigma} + \frac{q_\Sigma^2}{r_\Sigma^2}\right] = -\beta_{(r,t)} \frac{r_\Sigma^2 - 2\mu_\Sigma r_\Sigma + q_\Sigma^2}{r_\Sigma^2} \quad \text{(eq. C3)}$$

The Reissner-Nördstrom metric presents two event horizons: one outer horizon and one inner horizon, that accomplish the following condition:

$$\frac{dr_H}{dt} = 0 \Leftrightarrow r_H^2 - 2\mu r_H + q^2 = 0 \quad \text{(eq. C4)}$$

The solutions to equation "C4" are these ones:

$$\frac{dr_H}{dt} = 0 \Leftrightarrow r_H^2 - 2\mu r_H + q^2 = 0 \quad \text{(eq. C5a)}$$

$$r_H^{outer} = \mu + \sqrt{\mu^2 - q^2} \quad \text{(eq. C5d)}$$

$$r_H^{inner} = \mu - \sqrt{\mu^2 - q^2} \quad \text{(eq. C5c)}$$

In our model of collapse, we will assume that the collapsing energy-matter is nearing to form an outer horizon, so that the following change of variables will be useful:

$$r_\Sigma^\Delta = r_\Sigma - \left(\mu_\Sigma + \sqrt{\mu_\Sigma^2 - q_\Sigma^2}\right) \quad \text{(eq. C6)}$$

With this change of variables, equation "C3" may be rewritten in the following way:

$$\frac{dr_\Sigma^\Delta}{dt} = -\beta_{(r,t)} \frac{2\sqrt{\mu_\Sigma^2 - q_\Sigma^2}\, r_\Sigma^\Delta}{\left(\mu_\Sigma + \sqrt{\mu_\Sigma^2 - q_\Sigma^2}\right)^2} + O\left((r_\Sigma^\Delta)^2\right) \quad \text{(eq. C7)}$$

As we are evaluating the evolution of the system when $r_\Sigma^\Delta \ll r_\Sigma$, we are going to ignore all the $O\left((r_\Sigma^\Delta)^2\right)$ terms.

Therefore, the separation of variable becomes straightforward:

$$\frac{dr_\Sigma^\Delta}{r_\Sigma^\Delta} \approx -\beta_{(r,t)} \frac{2\sqrt{\mu_\Sigma^2 - q_\Sigma^2}}{\left(\mu_\Sigma + \sqrt{\mu_\Sigma^2 - q_\Sigma^2}\right)^2} dt \quad \text{(eq. C8)}$$

The solution to "C8" is the following one:

$$\ln r_\Sigma^\Delta \approx \frac{-2\sqrt{\mu_\Sigma^2 - q_\Sigma^2}}{\left(\mu_\Sigma + \sqrt{\mu_\Sigma^2 - q_\Sigma^2}\right)^2} \int \beta_{(r,t)} dt + A(\mu_\Sigma, q_\Sigma) \quad \text{(eq. C9a)}$$

$$r_\Sigma^\Delta \approx e^{A(\mu_\Sigma, q_\Sigma) - \frac{2\sqrt{\mu_\Sigma^2 - q_\Sigma^2}}{\left(\mu_\Sigma + \sqrt{\mu_\Sigma^2 - q_\Sigma^2}\right)^2} \int \beta_{(r,t)} dt} \quad \text{(eq. C9b)}$$

Now, in order to simplify the expressions, let's introduce a factor consisting in the quotient between the charge $q$ and the mass $\mu$ contained inside the Σ-surface:

$$\alpha_\Sigma \equiv \frac{q_\Sigma}{\mu_\Sigma} \quad \text{(eq. C10)}$$

With this parameter, we have:

$$r_\Sigma = \mu_\Sigma \left(1 + \sqrt{1 - \alpha_\Sigma^2}\right) + r_\Sigma^\Delta \approx \mu_\Sigma \left(1 + \sqrt{1 - \alpha_\Sigma^2}\right) \quad \text{(eq. C11a)}$$

$$\mu_\Sigma \approx \frac{r_\Sigma}{1 + \sqrt{1 - \alpha_\Sigma^2}} \quad \text{(eq. C11b)}$$

$$\ln r_\Sigma^\Delta \approx \frac{-2\sqrt{1 - \alpha_\Sigma^2}}{\mu_\Sigma \left(1 + \sqrt{1 - \alpha_\Sigma^2}\right)^2} \int \beta_{(r,t)} dt + A(\mu_\Sigma, q_\Sigma) \approx \frac{-2\sqrt{1 - \alpha_\Sigma^2}}{r_\Sigma \left(1 + \sqrt{1 - \alpha_\Sigma^2}\right)} \int \beta_{(r,t)} dt + A(r_\Sigma) \quad \text{(eq. C11c)},$$

where:

$$A(r_\Sigma) \equiv A(\mu(r_\Sigma), q(r_\Sigma)) \quad \text{(eq. C11d)}$$

Let's define another function, in order to be able tor write more compact equations:

$$F_\alpha(\alpha_\Sigma) \equiv \frac{2\sqrt{1 - \alpha_\Sigma^2}}{r_\Sigma \left(1 + \sqrt{1 - \alpha_\Sigma^2}\right)} \quad \text{(eq. C12)}$$

Then:

$$\ln r_\Sigma^\Delta \approx \frac{-F_\alpha(\alpha_\Sigma)}{r_\Sigma} \int \beta_{(r,t)} dt + A(r_\Sigma) \quad \text{(eq. C13a)}$$

$$1 - \frac{2\mu_\Sigma}{r_\Sigma} + \frac{q_\Sigma^2}{r_\Sigma^2} = \frac{r_\Sigma^2 - 2\mu_\Sigma r_\Sigma + q_\Sigma^2}{r_\Sigma^2} \approx \frac{2\sqrt{\mu_\Sigma^2 - q_\Sigma^2}}{\left(\mu_\Sigma + \sqrt{\mu_\Sigma^2 - q_\Sigma^2}\right)^2} r_\Sigma^\Delta \approx \frac{F_\alpha(\alpha_\Sigma)}{r_\Sigma} e^{A(r_\Sigma) - \frac{F_\alpha(\alpha_\Sigma)}{r_\Sigma} \int \beta_{(r,t)} dt} \quad \text{(eq. C13b)}$$

If it is taken into account that the previous expressions are valid for every Σ-surface, the Reissner-Nördstrom metric may be modified in order to describe the process of spherically symmetric collapse of a charged non rotating amount of energy-mass in the following way:

$$ds^2 = -h_{(r,t)} dt^2 + h_{(r,t)}^{-1} dr^2 + r^2 d\Omega^2 \quad \text{(eq. C14a)}$$

$$h_{(r,t)} = H(r - r_{edge}) h_{(r,t)}^{outside} + H(r_{edge} - r) h_{(r,t)}^{inside} \quad \text{(eq. C14b)}$$

$$h_{(r,t)}^{outside} = 1 - \frac{2M}{r} + \frac{Q^2}{r^2} \quad \text{(eq. C14c)}$$

$$h_{(r,t)}^{inside} = e^{f(r) - \frac{F_\alpha(\alpha_{(r)})}{r} \int \beta_{(r,t)} dt} \quad \text{(eq. C14d)}$$

$$\alpha_{(r)} \equiv \frac{q_{(r)}}{\mu_{(r)}} \quad \text{(eq. C14e)}$$

$$F_\alpha(\alpha_{(r)}) \equiv \frac{2\sqrt{1 - \alpha_{(r)}^2}}{r_\Sigma \left(1 + \sqrt{1 - \alpha_{(r)}}\right)}$$

$$r_{edge} \equiv \left(M + \sqrt{M^2 - Q^2}\right) + e^{A(r = M + \sqrt{M^2 - Q^2}) - \frac{2\sqrt{M^2 - Q^2}}{\left(M + \sqrt{M^2 - Q^2}\right)^2} \int \beta(r = M + \sqrt{M^2 - Q^2}, t) dt} \quad \text{(eq. C14g)}$$

$$f(r) \equiv A(r) + \ln\left(\frac{F_\alpha(\alpha(r))}{r}\right) \quad \text{(eq. C14h)}$$

The formulae are qualitatively very similar to the Never-Stationary Gravitational Collapse metric. A significative difference difference must be pointed out: to determine the temporal evolution of $\beta_{(r,t)}$ the electromagnetic force between shells should be considered (the model of collapse for initial states proposed in the previous section would not be properly applicable in this case), for instance by analysing the equations resulting from equaling the second covariant derivatives of coordinates to the components of the electromagnetic force.

**D) Some considerations about symmetry**

In reference "[3]" an argument is presented against Schwarzschild's black holes based in the symmetry of Schwarzschild metric under temporal inversion. If we replace $t$ by $-t$, the metric does not change, and therefore for any given trajectory the opposite trajectory should be possible as well. Thus, Schwarzschild metric should not be interpreted as describing a "black hole", but two isolated regions of space with no possible communication between them: the event horizon should not be understood as a "point without return", but as a surface incapable of being trespassed, not in one sense nor in the other one. On the other hand, the complete isolation from one region in respect to the other should make it completely impossible the apparition of the system in "history": the horizon should exist since the beginning of time, or it should never be formed at all.

All these considerations about the symmetry of Schwarzschild metric under temporal inversion are completely extensible to Reissner-Nördstrom metric. Notwithstanding, the two remaining types of black holes, Kerr's black hole and Kerr-Newman's black hole, are not symmetric under temporal inversion. The explanation to this fact is easily understandable: both Kerr and Kerr-Newman metrics describe rotating black holes. Given a trajectory of a particle falling inwards a rotating black hole, under temporal inversion not only the particle is travelling outwards, but the black hole is rotating in the opposite sense.

As Kerr metric consists in the particular case of Kerr-Newman metric when the charge of the rotating black hole is zero (as Schwarzshild and Reissner-Nördstrom metrics may be respectively understood as particular cases of Kerr and Kerr-Newman metric when there is no rotation), the whole problem can be dealt just by analysing Kerr-Newman metric, which is given by the following components:

$$ds^2 = \left[g_{tt}(r,\theta) dt^2 + g_{rr}(r,\theta) dr^2 + g_{\theta\theta}(r,\theta) d\theta^2 + g_{\phi\phi}(r,\theta) d\phi^2\right] + 2 g_{\phi t}(r,\theta) d\phi dt \quad \text{(eq. D1a)}$$

$$g_{tt}(r,\theta) = \frac{-\left[r^2 - 2Mr + \left(\frac{J}{M}\right)^2 \cos^2\theta + Q^2\right]}{r^2 + \left(\frac{J}{M}\right)^2 \cos^2\theta} \quad \text{(eq. D1b)}$$

$$g_{rr}(r,\theta) = \frac{r^2 + \left(\frac{J}{M}\right)^2 \cos^2\theta}{\left[r^2 - 2Mr + \left(\frac{J}{M}\right)^2 + Q^2\right]} \quad \text{(eq. D1c)}$$

$$g_{\theta\theta}(r,\theta) = r^2 + \left(\frac{J}{M}\right)^2 \cos^2\theta \quad \text{(eq. D1d)}$$

$$g_{\phi\phi}(r,\theta) = \frac{\sin^2\theta \left[\left(r^2 + \left(\frac{J}{M}\right)^2\right)^2 - \left(\frac{J}{M}\right)^2 \sin^2\theta \left(2Mr + \left(\frac{J}{M}\right)^2 + Q^2\right)\right]}{r^2 + \left(\frac{J}{M}\right)^2 \cos^2\theta} \quad \text{(eq. D1e)}$$

$$g_{\phi t}(r,\theta) = g_{t\phi}(r,\theta) = \frac{\frac{J}{M}(Q^2 - 2Mr)\sin^2\theta}{r^2 + \left(\frac{J}{M}\right)^2 \cos^2\theta} \quad \text{(eq. D1f)}$$

The crossed terms $g_{\phi t}(r,\theta)$ and $g_{t\phi}(r,\theta)$ are responsible of the temporal non-reversibility of the metric defined by equations "D1a-f". However, it may be noted that this metric is actually symmetric under simultaneous inversion of $t$ and $\phi$ :

$$ds^2(-t, r, \theta, -\phi) = ds^2(t, r, \theta, \phi) \quad \text{(eq. D2)}$$

Consequently, "looking backwards" in time implies a change in the system so that it appears rotating in the opposite sense, but if we correct this effect with a "mirror" (the inversion in $\phi$) the system becomes again completely identical to the original one. Thus, every particle which travels between from a radius $r_A$ to a radius $r_B$ should be capable of travelling backwards from the $r_B$ to the radius $r_A$, even when the trajectory may not considered to have been "reversed", for the system has kept on rotating meanwhile.

As a result, we should conclude that the same oppositions that we pointed out about the logical impossibility of formation of Schwarzschild's black holes can be equally applied to the more general type of black hole.

# Discussion and Conclusions

What is not logical cannot be true, but what is logical often may be true or not. Black holes, as they are commonly understood, contain a number of contradictory elements, and therefore it should be expected that they were finally rejected. In spite of that, there seems to be more than an alternative. On the one side, there exists the proposal of those who oppose the Hilbert's space-time as a correct concretion of Schwarzschild metric. On the other side there is the proposal of those who consider it as a good solution for the punctual mass, but who claim that the laws of Einstein's General Relativity prevent themselves from the formation of such structures. In this article, several aspects of the second hypothesis have been studied, and it has been shown how it is absolutely compatible with Einstein Field Equations. It is also consistent with the vision of gravitational collapse suggested by the findings of quantum emission of radiation performed by Vachaspati, Stojkovic and Krauss.

*Never Stationary Collapse* provides an explanation as good as black holes for every phenomenon in which the existence of black holes has been suggested to be implied, as the gravitational field outside the edge of the collapsing star coincides exactly with the one that a black hole would produce. The main difference dwells on the fact that its theoretical mathematical and physical basis is much more consistent, and directly deducible from the laws of Gravity. There is still much investigation to be performed in the field, and we concretely point out the need to search an adequate equation of state for collapsing matter, and the development of suitable a "never-stationary metric" for rotating collapse.

The only four exact solutions of General Relativity can be demonstrated to be impossible to be produced from any previous situation in which they were not already present. Nonetheless, that must not be a great drama: as far as we can calculate whatever we want with a degree of precision compatible with evidence, there is no need to ask anything more but logical coherence.

# Acknowledgements

We wish to thank Dr. Dejan Stojkovic for commenting to us our previous work and for making us to notice his exceedingly interesting papers.